\documentclass[final,5p,times,twocolumn]{elsarticle}

\usepackage[utf8]{inputenc}
\usepackage{amsmath,amssymb}
\usepackage{graphicx}
\usepackage{hyperref}
\usepackage{lineno}
\usepackage{comment}
\usepackage{xcolor}
\usepackage{textcomp}
\modulolinenumbers[5]

\journal{Astroparticle Physics}

\begin{document}

\begin{frontmatter}

\title{Monitoring the upper atmospheric temperature and interplanetary magnetic field with the GRAPES-3 muon telescope}

\author[tifr]{S. Paul}
\author[cust]{K.P. Arunbabu}
\author[tifr]{M. Chakraborty}
\author[tifr]{S.K. Gupta}
\author[tifr]{B. Hariharan\corref{cor1}} \ead{89hariharan@gmail.com} \cortext[cor1]{Corresponding author}
\author[ocu]{Y. Hayashi}
\author[tifr]{P. Jagadeesan}
\author[tifr]{A. Jain}
\author[tifr]{M. Karthik}
\author[cu]{H. Kojima}
\author[ocu]{S. Kawakami}
\author[tifr]{P.K. Mohanty}
\author[nu]{Y. Muraki}
\author[tifr]{P.K. Nayak}
\author[tu]{T. Nonaka}
\author[cu]{A. Oshima}
\author[tifr]{M. Rameez}
\author[tifr]{K. Ramesh}
\author[cu]{S. Shibata}
\author[hcu]{K. Tanaka}

\affiliation[tifr]{organization={Tata Institute of Fundamental Research},
            addressline={Dr. Homi Bhabha Road}, 
            city={Mumbai},
            postcode={400005}, 
            country={India}}

\affiliation[cust]{organization={Cochin University of Science and Technology},
            addressline={School of Environmental Studies},
            city={Kochi},
            postcode={682022}, 
            country={India}}

\affiliation[ocu]{organization={Osaka City University},
            addressline={Graduate School of Science}, 
            city={Osaka},
            postcode={558-8585}, 
            country={Japan}}

\affiliation[cu]{organization={Chubu University},
            addressline={College of Engineering}, 
            city={Chubu},
            postcode={487-8501}, 
            country={Japan}}

\affiliation[nu]{organization={Nagoya University},
            addressline={Institute for Space-Earth Environmental Research}, 
            city={Nagoya},
            postcode={464-8601}, 
            country={Japan}}

\affiliation[tu]{organization={University of Tokyo},
            addressline={Institute for Cosmic Ray Research}, 
            city={Tokyo},
            postcode={277-8582}, 
            country={Japan}}

\affiliation[hcu]{organization={Hiroshima City University},
            addressline={Graduate School of Information Sciences}, 
            city={Hiroshima},
            postcode={731-3166}, 
            country={Japan}}

\begin{abstract}
    We study the influence of variations in the upper atmospheric temperature and interplanetary magnetic field on the cosmic ray induced atmospheric muon flux measured by the GRAPES-3 experiment over 22 years (2001--2022) of data; spanning three solar cycles: the declining phase of Solar Cycle 23, the full Cycle 24, and the rising and maximum phases of Cycle 25. Located in Ooty- India, the GRAPES-3 large area ($560\,\mathrm{m}^2$) muon telescope detects $\sim$4 billion muons daily above $1\,\mathrm{GeV}$, with an angular resolution of $\sim$4$^\circ$, enabling a statistical precision $<$0.01\% on the hourly muon rate. After accounting for the effect of atmospheric pressure variations, we compare this data with the upper atmospheric temperature inferred from NASA’s MERRA-2 dataset as well as magnetic field data from the ACE and WIND spacecraft at Lagrange point L1. A simultaneous iterative fitting method employing Fast Fourier Transforms and a narrow band-pass filter reveals the temperature and magnetic field coefficients to be $\alpha_{\text{T}}=-\,0.2241\,\pm\,0.04\, (\text{stat.})\,\pm\,0.0220\, (\text{syst.})\,\%\,\text{K}^{-1}$ and $\gamma_{\text{M}}=-\,0.574\,\pm\,0.027\, (\text{stat.})\,\pm\,0.011\, (\text{syst.})\,\%\,\text{nT}^{-1}$, respectively, for an assumed hadronic attenuation length $\lambda$=120 g cm$^{-2}$, underscoring the potential of the GRAPES-3 muon telescope to serve as a real time monitor of the upper atmospheric temperature or interplanetary magnetic field.
    
\end{abstract}


\end{frontmatter}


\section{Introduction}

Galactic Cosmic Rays (GCRs) have to travel through the heliosphere before they interact with the Earth's atmosphere. During this, they are deflected by the Sun's magnetic field, causing variations in this field to imprint on the flux, spectrum and angular distribution of GCRs detected at or near Earth. Studies of these variations over the past several decades have revealed the impact of both transient phenomena such as solar flares, coronal holes, sunspot activity and coronal mass ejections (CMEs) as well as their effects such as Forbush Decreases (FDs), precursors and Ground-Level Enhancements (GLEs)\cite{1954JGR,2015SoPh,2015AA}. Periodic variations, such as due to the solar diurnal modulation, the 27-day solar rotation, the 11-year solar cycle, and the 22-year solar magnetic cycle \cite{2013Prama..81..343M, 1990SSRv...52..121V} have also been characterized. These Sun-induced phenomena are most prominent in GCR intensity variations up to $\sim$30\,GeV/nuc, beyond which the influence of solar modulation decreases rapidly \cite{2013LRSP...10....3P} as the gyro-radii of GCRs exceed the characteristic size of the heliosphere ($\sim$100\,AU).

Cosmic rays are high-energy particles ($\sim$10$^{9}\,\mathrm{eV} \text{--}10^{20}\,\mathrm{eV}$
), primarily protons ($\sim$90\%) and alpha particles ($\sim$9\%), with a small fraction of heavier nuclei ($\sim$1\%) and trace amounts of positrons and antiprotons \cite{2016crpp.book.....G}. When they enter the Earth's atmosphere, they collide with nuclei, mainly Oxygen and Nitrogen, and produce an increasing flux of secondary particles while propagating downwards \cite{1960ARNPS..10...63G}. The secondary particles in the upper atmosphere are largely neutrons and charged mesons (mainly pions and kaons). Due to their short decay lengths ($\sim$550\,m for 10\,GeV charged pions), these mesons decay into muons and neutrinos. Most muons survive to reach Earth’s surface due to their relatively long decay lengths, with high-energy muons even penetrating deep underground, while neutrinos typically traverse the Earth without interaction. 

While traveling downwards in the atmosphere, muons lose energy via ionization. Most atmospheric muons are produced at altitudes $>$10\,km and lose $\sim$2\,GeV of energy before reaching the ground. The mean energy of muons near sea level has been measured to be $\sim$4\,GeV, and their energy spectrum is flat up to 1\,GeV, gradually steepening at higher energies~\cite{2004hep.ex....8114T}. The integral flux of $>$1\,GeV muons at the sea level is $\sim$70\,$\mathrm{m}^{-2}\mathrm{s}^{-1}\mathrm{sr}^{-1}$\cite{2022PTEP.2022h3C01W}. The angular distribution of the muon varies as $\cos^{2}\theta$
, where $\theta$ is the zenith angle \cite{2001AdSpR..28.1773S}.


The \textbf{G}amma \textbf{R}ay \textbf{A}stronomy at \textbf{P}eV \textbf{E}nergie\textbf{S} - phase \textbf{3} (GRAPES-3) experiment \cite{2005NIMPA.545..643H} is a ground-based air shower observatory in Ooty, India. The large area muon telescope of GRAPES-3 has been measuring the flux of atmospheric muons with energies above 1 GeV, produced by the interactions of GCRs with the Earth’s atmosphere, over the last two solar cycles. The upper atmospheric temperature exhibits seasonal fluctuations caused by Earth's orbital motion around the Sun, leading to corresponding changes in atmospheric density. Consequently, the muon flux also shows a seasonal variation correlated with these temperature variations. This seasonal variation was first postulated in 1952, and a fairly comprehensive phenomenological framework of its explanation was also outlined \cite{1952RvMP...24..133B}. GRAPES-3 detects low-energy muons produced by mesons that decay almost entirely at high altitudes. At GRAPES-3's equatorial location, temperature variations in this region are minimal, making the influence of the upper atmospheric temperature on meson decay negligible. The resulting muons have energies of a few GeV, and because they typically lose $\sim$2\,GeV before reaching the ground, a fraction of them decay during propagation. An increase in the upper atmospheric temperature causes an outward expansion of the atmosphere, which in turn increases the path length traveled by the muons. As a result, the probability of muon decay increases. Consequently, low-energy muons (a few GeV) show a negative dependence on the upper atmospheric temperature.

In contrast, underground experiments like MACRO \cite{1997APh.....7..109M}, MINOS \cite{2008ICRC....5.1233G}, and IceCube \cite{2011ICRC....1...78D} detect muons with much higher threshold energies ($\sim$100\,GeV), which originate in the decay of parent mesons of commensurately higher energies. An increase in the upper atmospheric temperature causes atmospheric expansion, which increases the probability of meson decay. This process enhances the production of secondary muons. Additionally, the higher energy muons have lower probabilities of decaying; thus, almost all survive to reach the deep underground locations of these experiments. Consequently, high-energy muons display a positive dependence on the upper atmospheric temperature, which was measured by the MINOS and IceCube detectors, due to a large seasonal variation in the temperature of the atmosphere($\sim$10\,K) at their high-latitude locations \cite{2011ICRC....1...78D, 2014PhRvD..90a2010A}.
 
Apart from seasonal variations, daily fluctuations in upper atmospheric temperature also affect the decay rate of muons produced by GCRs, which modulates the detected muon flux. Additionally, Earth’s rotation within the ambient solar wind causes a solar diurnal variation in muon detection rates. This makes it difficult to disentangle these two phenomena responsible for daily muon rate variations. To address this, seasonal variations in upper atmospheric temperature were used to isolate their effects on muon flux variations observed by the GRAPES-3 muon telescope \cite{2017APh....94...22A}. While the earlier work confined itself to six years (2005--2010) of data. In this work, we extend this analysis to twenty-two years (2001--2022) of data. Since the variation in the interplanetary magnetic field over this period spanning three Solar cycles is substantial, we iteratively deconvolute its effects from that of the upper atmospheric temperature by fitting for the linear coefficients of both simultaneously. The resulting best fit parameterization enables the muon flux recorded by the GRAPES-3 muon telescope to be used as a relative measure of the upper atmospheric temperature or the  magnetic field at L1 to within 20\% and 6\%, respectively. 

This paper is organized as follows. In Section 2 we describe the GRAPES-3 observatory, with a focus on the muon telescope, which collected the data forming the primary focus of this paper. Section 3 describes the datasets and the processing applied to them before analysis. Section 4 presents the methodology utilizing Fast Fourier Transforms to iteratively fit for the temperature coefficient ($\alpha_{\text{T}}$) and the magnetic field coefficient ($\gamma_{\text{M}}$) simultaneously. In section 5 we summarize the results, which we discuss and conclude in section 6.


\section{The GRAPES-3 observatory}

GRAPES-3 is a ground-based air shower observatory located in Ooty, South India ($11.4^\circ$ N, $76.7^\circ$ E) at an altitude of 2200 meters above mean sea level. It consists of two major components. The first is an extensive air shower (EAS) array of 400 plastic scintillator detectors \cite{2005NIMPA.540..311G}, arranged in a symmetric hexagonal configuration. Each detector has an area of $1$\,$\mathrm{m}^2$, with an inter-detector separation of 8 meters. The array was designed to study primary cosmic rays in the energy range of 10 TeV to 10 PeV and to perform precision measurements of their energy spectrum and composition. The second component is a large-area (560\,$\mathrm{m}^2$) muon telescope \cite{2005NIMPA.545..643H}, which continuously samples the atmospheric muon flux with a special emphasis on coincidence with EAS events observed by the scintillator array. 

\begin{figure}
    \centering
    \includegraphics[width=0.4\textwidth]{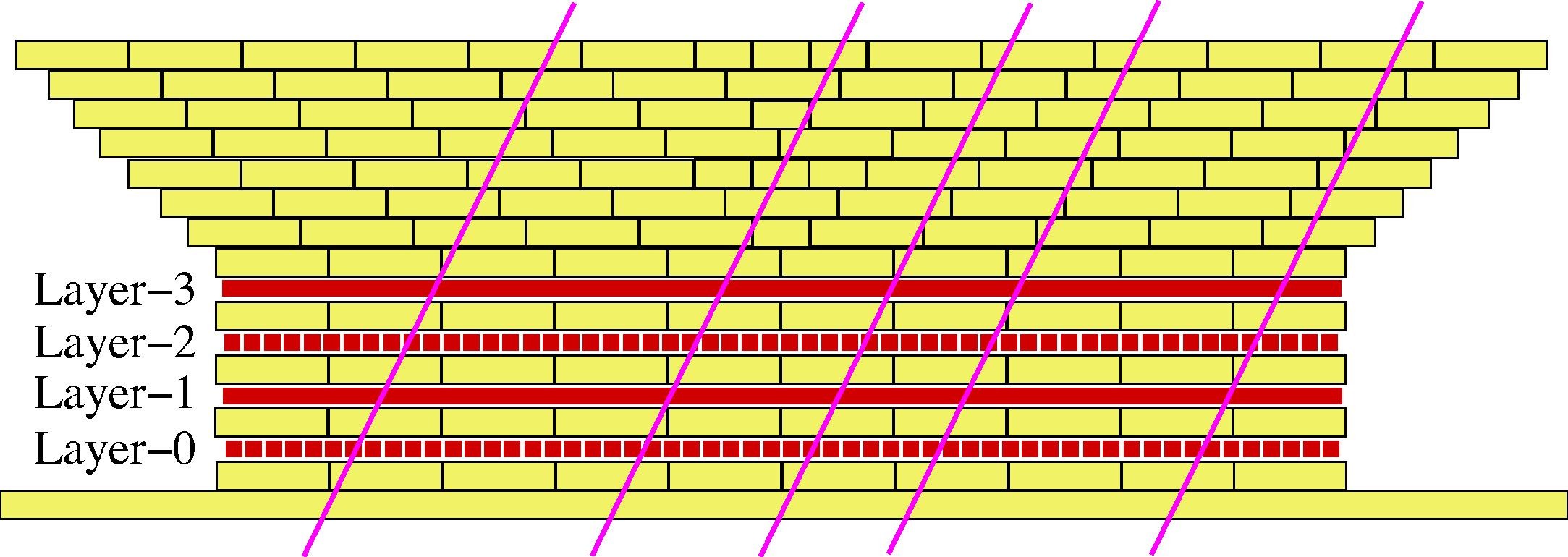}
    \caption{A schematic of a 4-layer tracking muon telescope with 58 PRCs per layer.}
    \label{Fig_01}
\end{figure}

The GRAPES-3 muon telescope consists of 16 independent modules, each covering an area of $35\,\mathrm{m}^2$. These modules are organized into four logical groups called super-modules. The basic element of the muon telescope is the proportional counter (PRC), a gas ionization detector constructed from a mild steel tube with dimensions 6 m $\times$ 10 cm $\times$ 10 cm and a wall thickness of 2.3 mm. The tubes are sealed at both ends and filled with a P10 gas mixture ($90\,\%$ argon and $10\,\%$ methane). A $100\,\mu\mathrm{m}$ tungsten wire runs through the center as the anode, while the tube body serves as the cathode. Each muon module contains 232 PRCs arranged in four layers of 58 PRCs each, with alternate layers aligned in mutually orthogonal directions (Figure \ref{Fig_01}). Successive PRC layers are separated by 15 cm of reinforced concrete slabs. A concrete block arrangement ($\sim$550\,$\mathrm{g\,cm}^{-2}$) shaped like an inverted pyramid provides a 1 GeV energy threshold for vertical muons, scaling as $\sec\theta$\,GeV for muons incident at zenith angle $\theta$ (with coverage up to 45$^\circ$). Each of the 16 modules in the telescope was instrumented to provide a continuous record of the muon flux in 15$\times$15 directional bins, covering a solid angle of 2.3\,sr at 10-second intervals. In the past, the high-statistics GRAPES-3 muon data have yielded path breaking insights into solar \cite{PhysRevLett.117.171101} and atmospheric \cite{Hariharan:2019fai} phenomena.

\section{The datasets}

\begin{figure*}
    \centering
    \includegraphics[width=\textwidth]{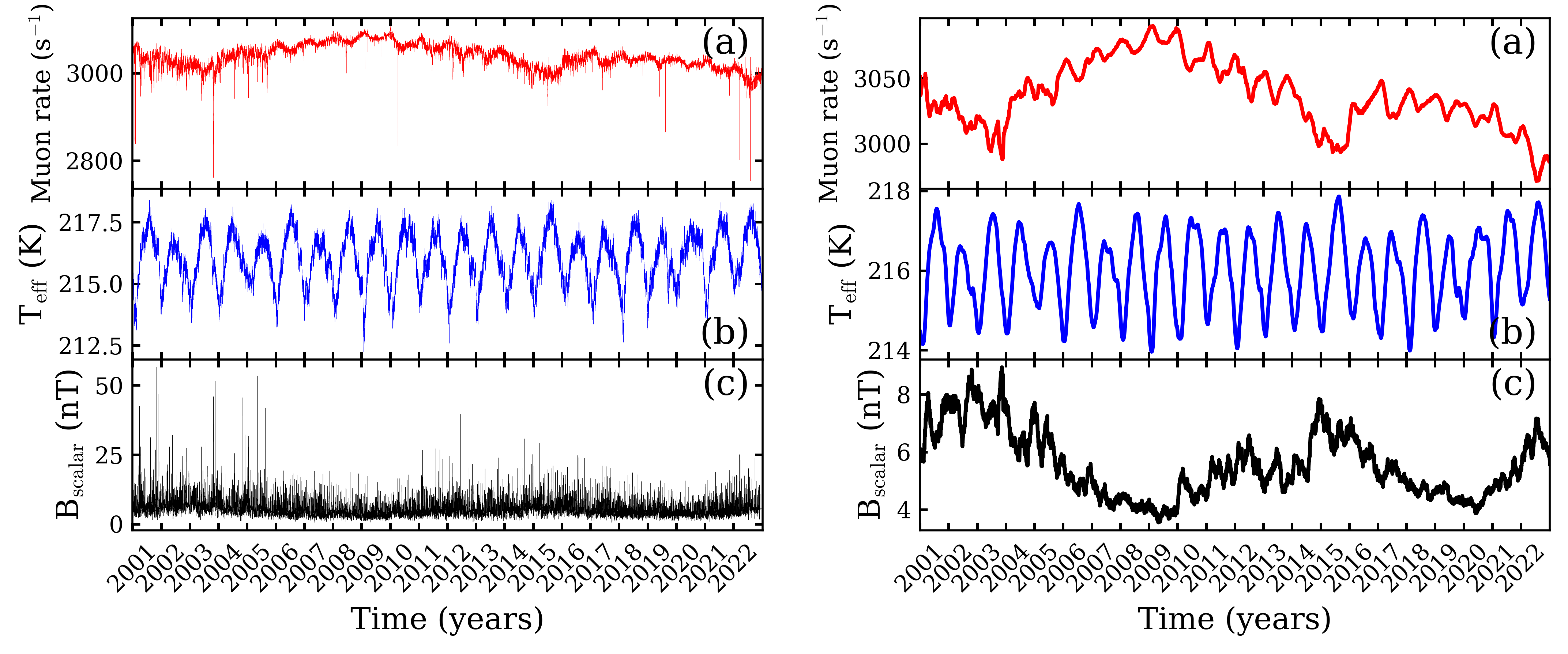}
    
   \caption{Temporal variations of (a) the muon rate, (b) the effective temperature (T$_\text{eff}$; assuming $\lambda$=120 g\,cm$^{-2}$), and (c) the interplanetary magnetic field $(\text{B}_{\text{scalar}}=\sqrt{\text{B}_{x}^{2}+\text{B}_{y}^{2}+\text{B}_{z}^{2}})$ observed by the ACE and WIND spacecraft \cite{1998SSRv...86..613S,1995SSRv...71..207L,2005JGRA..110.2104K} at Lagrange point L1, during 22 years (2001--2022). The left panels show the 3-hourly averaged data, while the right panels present the same datasets after applying a 60-day running average to suppress short term variations.}
    \label{Fig_3}
\end{figure*}

This analysis utilizes 22 years (2001--2022) of continuous muon flux measurements recorded by the GRAPES-3 muon telescope (G3MT) at Ooty, India, spanning over three solar cycles: the declining phase of Solar Cycle 23, the full Cycle 24, and the rising and maximum phases of Cycle 25. The G3MT  modules continuously record muon rates in 225 angular directions, in time bins of ten second duration and an angular coverage of 2.3\,sr. These high-statistics data enable statistical uncertainties as low as $\sim$0.002\% for daily averaged muon rates and $\sim$0.06\% for minute-scale rates, allowing the detection of subtle seasonal and solar-induced modulations. Despite careful maintenance and operations, instrumental effects arising from momentary or localized issues in signal processing and gradual changes in detector efficiency have been observed. For the first time, an automated correction algorithm based on Bayesian Blocks and the Savitzky-Golay filter, developed and validated in \cite{Paul:2025iet}, was implemented to mitigate all identified instrumental effects, ensuring long-term detector stability within the data. This algorithm capitalizes on the redundancy offered by the sixteen independent modules of G3MT to distinguish true variations of interest from variations in detector efficiency. The efficiency-corrected muon rate is summarized in the left panel of
Figure~\ref{Fig_3}(a).

Atmospheric pressure is measured at the GRAPES-3 site every calendar minute using Vaisala PTB200 and PTB220 digital barometers \cite{Vaisala}. These barometers are placed inside Station-1 of the GRAPES-3 muon telescope, where the temperature and humidity are maintained under stable conditions. The pressure effect on muon flux is corrected using a pressure coefficient ($\beta$) = --0.128\,\%$\mathrm{hPa}^{-1}$ \cite{2016APh....79...23M}. Subsequently we derive 3 hour averages of the muon data, downgrading the temporal resolution to match that of the upper atmospheric temperature data; leading to a typical statistical uncertainty much less than 0.01\,\%. It is not possible to determine the specific altitude at which individual muons originate. Thus to study the correlation between muon flux and atmospheric temperature, the atmosphere is modeled using an effective temperature, T$_{\text{eff}}$, which represents a weighted average of temperatures throughout the atmospheric column down to the observation level. Following Ref.~\cite{2017APh....94...22A} we estimate T$_{\text{eff}}$ using the expression: \begin{equation}
\text{T}_{\text{eff}}=\frac{\sum \exp{(\frac{-\text{x}}{\lambda})} \text{T(x)}\Delta \text{x}}{\sum \exp{(\frac{-\text{x}}{\lambda})}\Delta \text{x}},
\label{eq:Teff}
\end{equation} 
 where `x' is the atmospheric depth, in terms of the amount of matter above a point in the atmosphere, $\lambda$ is the hadronic attenuation length, which determines the depth at which hadrons interact before decaying, and T(x) is the temperature at depth x. The atmospheric depth x is linearly related to the atmospheric pressure $p$, with x (in g cm$^{-2}$) $\simeq$ 1.0195 $p$ (in hPa) \cite{2011ACP....11.1979U}. The hadronic attenuation length ($\lambda$) varies between 80 to 180 g\,cm$^{-2}$ depending on the particle type and interaction energy. Typical values are $\lambda_{\pi}\approx$ 160 g cm$^{-2}$ for pions and $\lambda_{N}\approx$ 120 g cm$^{-2}$ for nucleons \cite{2016crpp.book.....G,1952RvMP...24..133B,1997APh.....7..109M,2014PhRvD..90a2010A}. In this analysis, we adopt a central value of $\lambda = 120$ g cm$^{-2}$, and subsequently generalize the results to different values of $\lambda$.

\begin{figure}
    \centering
    \includegraphics[width=0.48\textwidth]{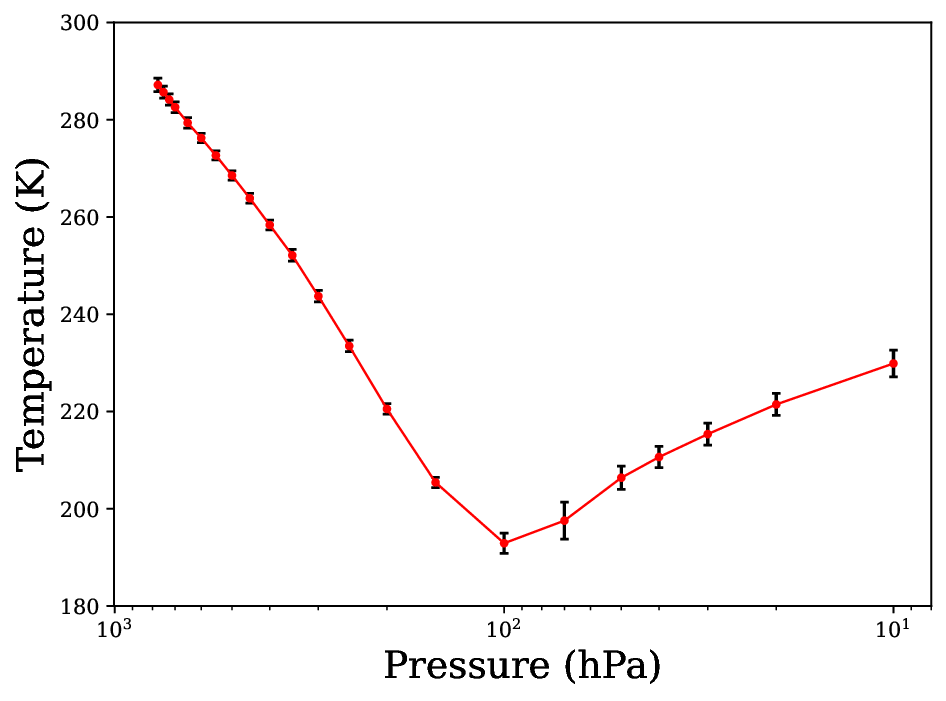}
    \caption{The 22-year average vertical temperature profile obtained from NASA's MERRA-2 dataset at the location ($\theta$ = 11.5$^\circ$, $\phi$ = 76.825$^\circ$) near the GRAPES-3 site, shown by the red line, as a function of pressure level. Temperatures are presented at 22 different pressure levels ranging from 775 hPa to 10 hPa above sea level. {\color{black} The error bars represent the standard deviation ($\sigma$) of the temperature, computed over the 22-year dataset.}}
    \label{Fig_1}
\end{figure}

We use temperature data from NASA’s MERRA-2 dataset \cite{GMAO_MERRA2_inst3_3d_asm_Np} for the same period as the muon flux data. {\color{black}Previous studies report a 15-year mean root-mean-square error (RMSE) of 0.962 K in the weighted mean temperature derived from MERRA-2 reanalysis data \cite{2022RemS...14.5431L}.} Temperature data covering a range of latitudes and longitudes near the GRAPES-3 site were used to estimate the effective temperature, T$_{\text{eff}}$, above Ooty. It has been shown that muons detected by GRAPES-3 are produced at altitudes between 6 km and 30 km above sea level \cite{2017APh....94...22A}. In this study, temperature data were obtained at 22 pressure levels (775, 750, 725, 700, 650, 600, 550, 500, 450, 400, 350, 300, 250, 200, 150, 100, 70, 50, 40, 30, 20, and 10 hPa), spanning from near the surface (775 hPa) to higher altitudes (10 hPa). Measurements were taken eight times daily at 00:00, 03:00, 06:00, 09:00, 12:00, 15:00, 18:00, and 21:00. Figure \ref{Fig_1} shows the 22-year average vertical temperature profile at the location ($\theta$ = 11.5$^\circ$, $\phi$ = 76.825$^\circ$) near the GRAPES-3 site, plotted as a function of pressure. {\color{black}The statistical uncertainty at each pressure level is estimated as the standard deviation ($\sigma$), calculated from the complete 22-year dataset, thus also including the seasonal variation.}

\begin{figure}
    \centering
    \includegraphics[width=0.48\textwidth]{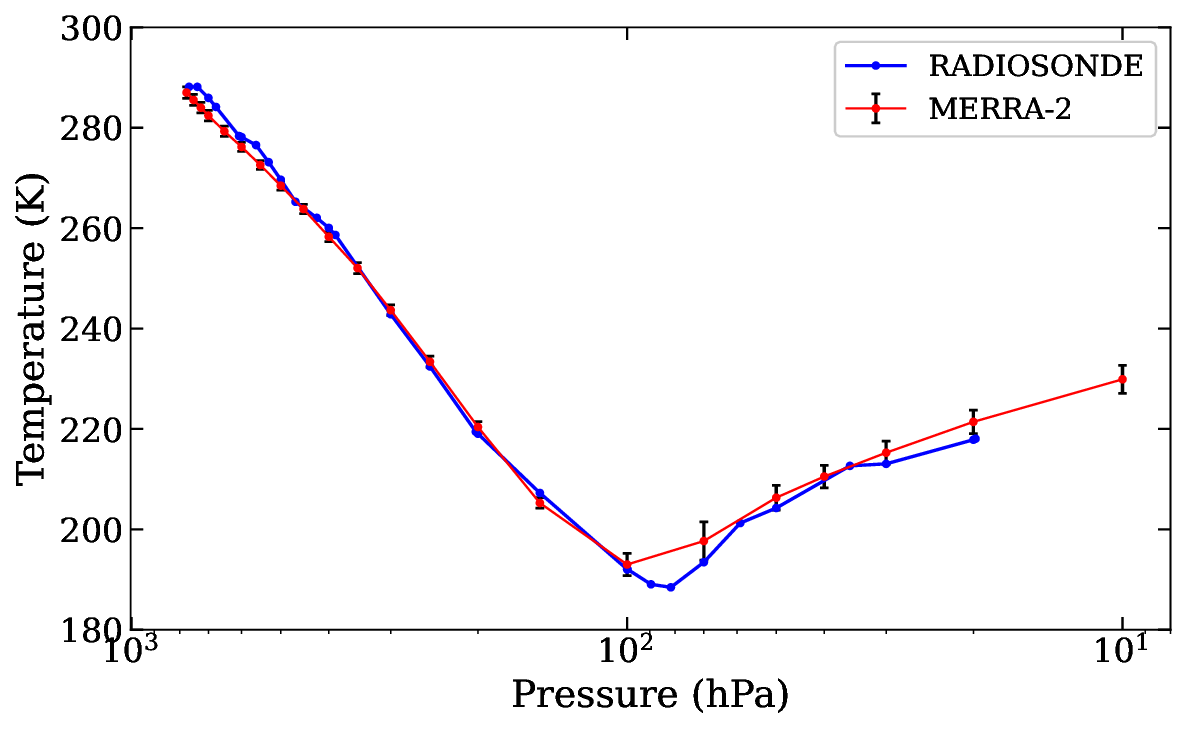}
    \caption{{\color{black}Vertical temperature profiles derived from MERRA-2 (red) and radiosonde observations (blue). The radiosonde data are from Cochin (9.95$^\circ$ N, 76.26$^\circ$ E) at 00:00 UTC on 2 January 2001, while the MERRA-2 profile represents the 22-year mean temperature profile corresponding to a nearby grid point (10.0$^\circ$ N, 76.25$^\circ$ E), located approximately 5.7 km from the radiosonde station.}}
    \label{Fig_1_1}
\end{figure}

The temperature at the exact GRAPES-3 site (11.4$^\circ$ N, 76.7$^\circ$ E), was obtained using a bilinear interpolation, performed using temperature data from four nearby grid points: ($\theta_1$ = 11.0$^\circ$, $\phi_1$ = 76.25$^\circ$), ($\theta_1$ = 11.0$^\circ$, $\phi_2$ = 76.825$^\circ$), ($\theta_2$ = 11.5$^\circ$, $\phi_1$ = 76.25$^\circ$), and ($\theta_2$ = 11.5$^\circ$, $\phi_2$ = 76.825$^\circ$), following the equation:
\begin{align}
\begin{split}
\text{T}(\theta, \phi) &= \frac{1}{[\theta_2 - \theta_1][\phi_2 - \phi_1]} \Big[ \text{T}(\theta_1, \phi_1)[\theta_2 - \theta][\phi_2 - \phi] + \\& \text{T}(\theta_2, \phi_1)[\theta - \theta_1)][\phi_2 - \phi]+ \text{T}(\theta_1, \phi_2)[\theta_2 - \theta][\phi - \phi_1] + \\&\text{T}(\theta_2, \phi_2)[\theta - \theta_1][\phi - \phi_1] \Big]
\end{split}
\end{align}
The left panel of Figure \ref{Fig_3}(b) shows the variation of the effective temperature ($\text{T}_{\text{eff}}$) at the GRAPES-3 site over 22 years (2001--2022).

{\color{black}To assess the reliability of the MERRA-2-derived effective temperature, we compared the vertical temperature profile from MERRA-2 with radiosonde observations. The radiosonde data were obtained from Cochin (9.95$^\circ$ N, 76.26$^\circ$ E) from the University of Wyoming (accessed on 16 July 2022) \cite{uwyo_radiosonde} at 00:00 UTC on 2 January 2001, while the MERRA-2 profile represents the 22-year mean temperature profile corresponds to the nearest grid point (10.0$^\circ$ N, 76.25$^\circ$ E), located approximately 5.7 km from the radiosonde station. As shown in Fig. \ref{Fig_1_1}, the two profiles show overall consistency across the atmospheric column, with the radiosonde temperatures lying within the $3\sigma$ uncertainty range of the MERRA-2 values at all pressure levels. This consistency supports the reliability of the MERRA-2-based effective temperature used in this study.}

In addition to atmospheric temperature-driven seasonal modulations, the muon rate exhibits long-term variations caused by the modulation of the primary CR flux by solar and heliospheric processes. To evaluate the influence of long-term solar modulation, we used interplanetary magnetic field (IMF) measurements from the NASA OMNI database. The OMNI provides magnetic field measurements from the ACE MAG and WIND MFI instruments located at the Lagrange point L1. These instruments provide magnetic field data, with a precision of 0.025$\%$ for both ACE and WIND and an absolute accuracy of 0.1 nT for ACE and $<$ 0.08 nT for WIND \cite{1998SSRv...86..613S,1995SSRv...71..207L}. In this work, we utilize the hourly-averaged IMF magnitude from OMNI, which was subsequently averaged to a 3-hour resolution to match the temporal binning of the muon and temperature series, as presented in the left panel of Figure \ref{Fig_3}(c). 

Since the muon flux exhibits modulations across multiple short term time scales in addition to the annual seasonal modulation caused by atmospheric temperature changes, a 60-day running average was applied to all three time series before further analysis. The data at this stage are presented in the right panels of Figure \ref{Fig_3}.


\section{Methodology}

\subsection{The temperature coefficient ($\alpha_{\text{T}}$)}
\label{section:temp_coeff}

\begin{figure}
    \centering
    \includegraphics[width=0.48\textwidth]{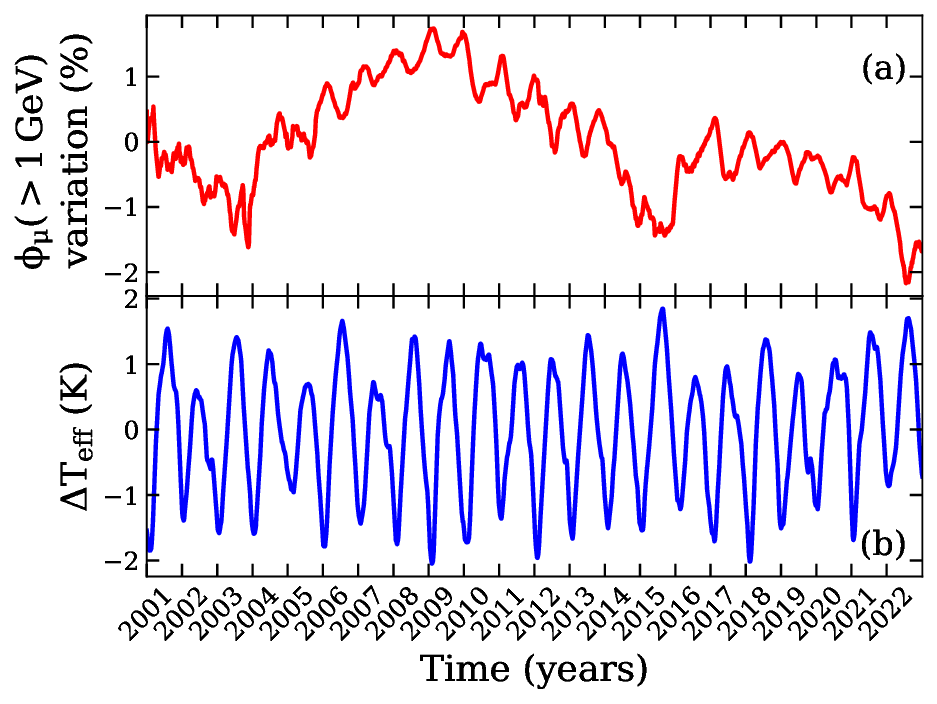}
    \caption{Variation of (a) muon flux (in $\%$), and (b) $\Delta{\text{T}_{\text{eff}}}$ (in K; assuming $\lambda$=120 g\,cm$^{-2}$), obtained using a 60-day low-pass filter (implemented via a running average) during 22 years (2001--2022).}
    \label{Fig_4}
\end{figure}

The dependence of the muon rate on atmospheric temperature is characterized by the temperature coefficient $\alpha_{\text{T}}$, which quantifies the percentage fractional change in muon intensity per unit change in effective atmospheric temperature. It is defined as:
\begin{equation}
    \frac{\text{R}-\Bar{\text{R}}}{\Bar{\text{R}}}\times 100(\%)=\alpha_{\text{T}}\times \Delta{\text{T}_{\text{eff}}},\label{eq:alpha_T}
\end{equation}
where R is the muon rate at a given time, $\Bar{\text{R}}$ represents the mean muon rate, and $\Delta{\text{T}_{\text{eff}}}$ = $\text{T}_{\text{eff}}-\Bar{\text{T}}_{\text{eff}}$ is the deviation of the effective temperature from its mean value. The chosen window length suppresses short-term variations while preserving both the slower seasonal modulations and the even longer-term solar cycle trends. Figures \ref{Fig_4}(a) and \ref{Fig_4}(b) display the percentage deviations in the muon flux and in $\Delta{\text{T}_{\text{eff}}}$ over the 22-year period (2001--2022) at this stage. A pronounced anti-correlation between muon flux variations and temperature deviations is observed, especially during solar minimum periods (e.g., 2008--2009 and 2019--2020), when heliospheric modulation is minimal and atmospheric effects dominate. In contrast, during solar maxima, this correlation weakens as heliospheric modulation becomes stronger, partially masking the temperature-driven variations in the muon flux.

\begin{figure}
    \centering
    \includegraphics[width=0.48\textwidth]{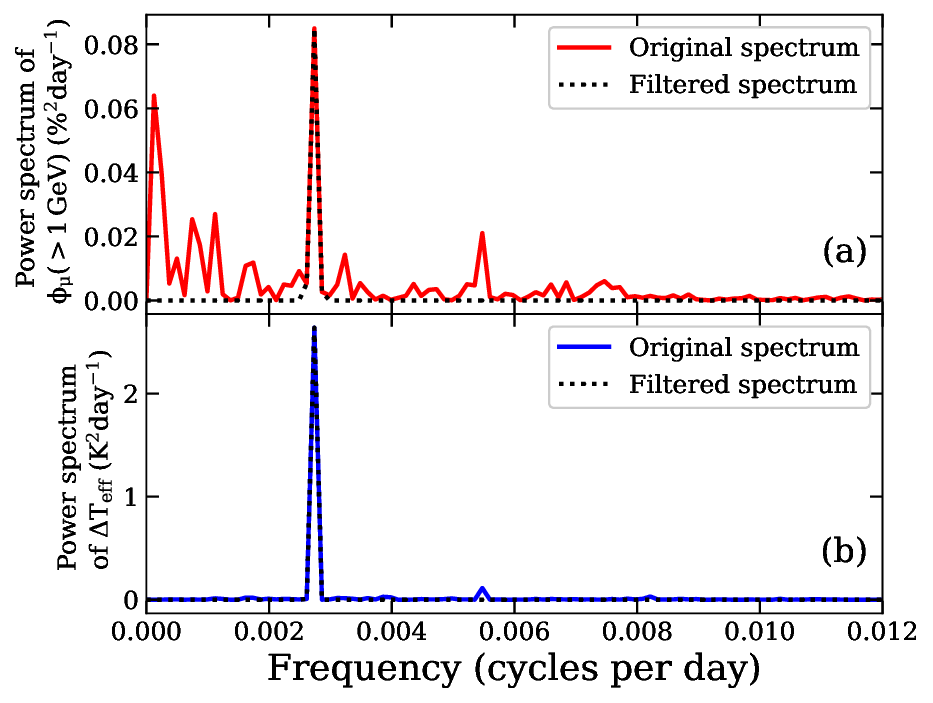}
    \caption{Fast Fourier transform (FFT) spectra of (a) muon flux variation (in $\%$) and (b) $\Delta{\text{T}_{\text{eff}}}$ (in K; assuming $\lambda$=120 g\,cm$^{-2}$) over the 22 years (2001--2022). Red and blue solid lines represent the original spectra for panels (a) and (b), respectively, while black dotted lines indicate the corresponding filtered spectra.}
    \label{Fig_5}
\end{figure}

Although low-pass filtering effectively suppresses short-term variations, it cannot isolate the annual modulation induced by atmospheric temperature from the slower 11-year solar cycle variations. To overcome this limitation, we employ a frequency-domain analysis based on the fast Fourier transform (FFT), which decomposes the time series into its constituent periodicities, enabling isolation of the annual temperature-driven component. The FFT analysis was implemented using the \texttt{numpy.fft.fft} function in Python. The resulting FFT spectra for the muon flux variation and temperature deviation are shown in Figures \ref{Fig_5}(a) and \ref{Fig_5}(b) as solid red and blue lines, respectively. Both spectra exhibit a distinct and coincident peak at 0.002738 cycles per day (CPD); corresponding to a 1-year periodicity, confirming that the seasonal modulation in muon flux originates from atmospheric temperature variations. This identified annual periodicity is then isolated using a narrow band-pass filter to suppress all non-atmospheric (solar-induced) frequencies, to enable a more accurate determination of $\alpha_{\text{T}}$. 

\begin{figure}
    \centering
    \includegraphics[width=0.48\textwidth]{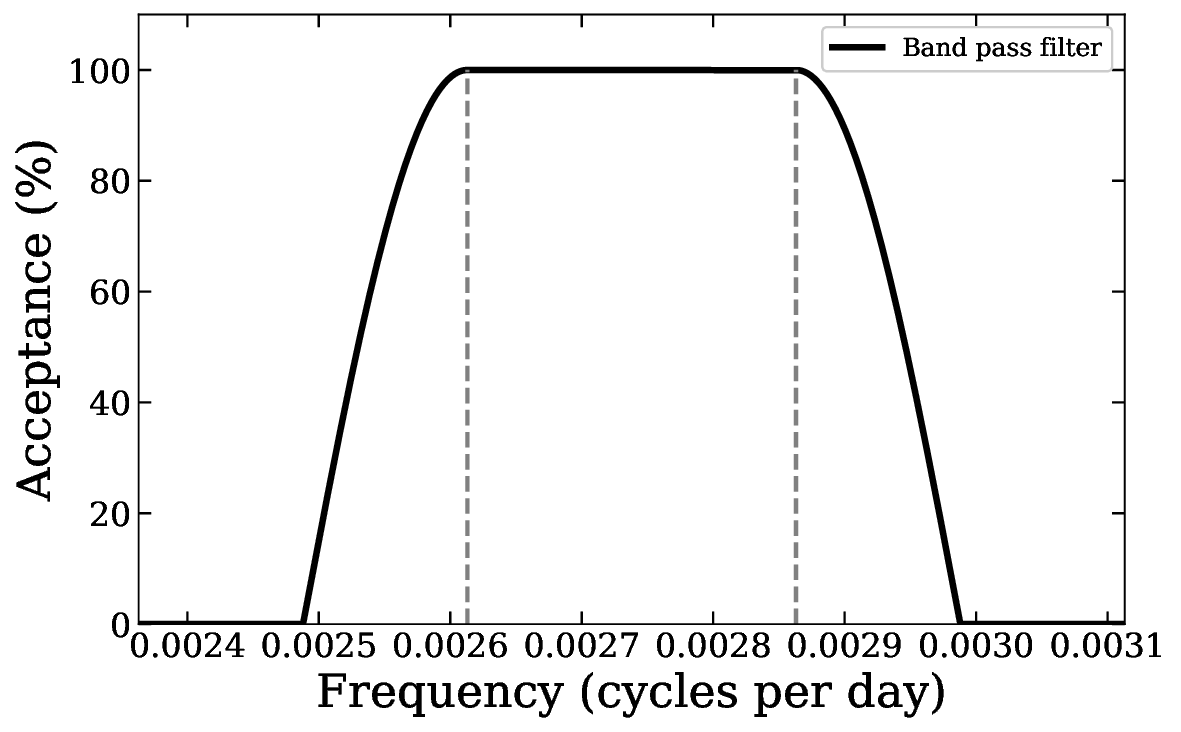}
    \caption{A band-pass filter was applied to the muon and temperature data to select frequencies centered at 0.002738 cycles per day. The two dashed vertical lines indicate the region of 100 $\%$ acceptance.}
    \label{Fig_6}
\end{figure}

The filter, incorporating a smooth sine transition at the edges to avoid spectral leakage and ensure continuity in the reconstructed signal, is defined as
 \begin{equation}
       W(f)=
        \begin{cases}
        1, & \text{if $|f-f_{C}|\leq \Delta f$} \\
        \sin{\frac{\pi}{2}\frac{|f-f_{C}|}{\Delta f}}, & \text{if $\Delta f<|f-f_{C}|\leq 2\Delta f$}\\
        0, & \text{if $|f-f_{C}|>2\Delta f$}.
        \end{cases}\label{eq:band-pass}
    \end{equation}
and is centered at $f_{C}$=0.002738 CPD with a half-width of $\Delta f$ = 0.000125 CPD (Figure \ref{Fig_6}). The half-width is chosen as $\Delta f=1/T$, where $T$ is the total duration of the dataset, ensuring that the filter bandwidth corresponds to the minimum frequency resolution achievable in the FFT spectrum. The filter maintains full acceptance between 0.002613 and 0.002863 CPD, tapering smoothly to zero beyond these limits to suppress adjacent frequencies. The resulting filtered spectra, shown as black dotted lines in Figures \ref{Fig_5}(a) and \ref{Fig_5}(b), clearly isolate the annual seasonal component in both the muon flux and temperature variation.

\begin{figure}
    \centering
    \includegraphics[width=0.48\textwidth]{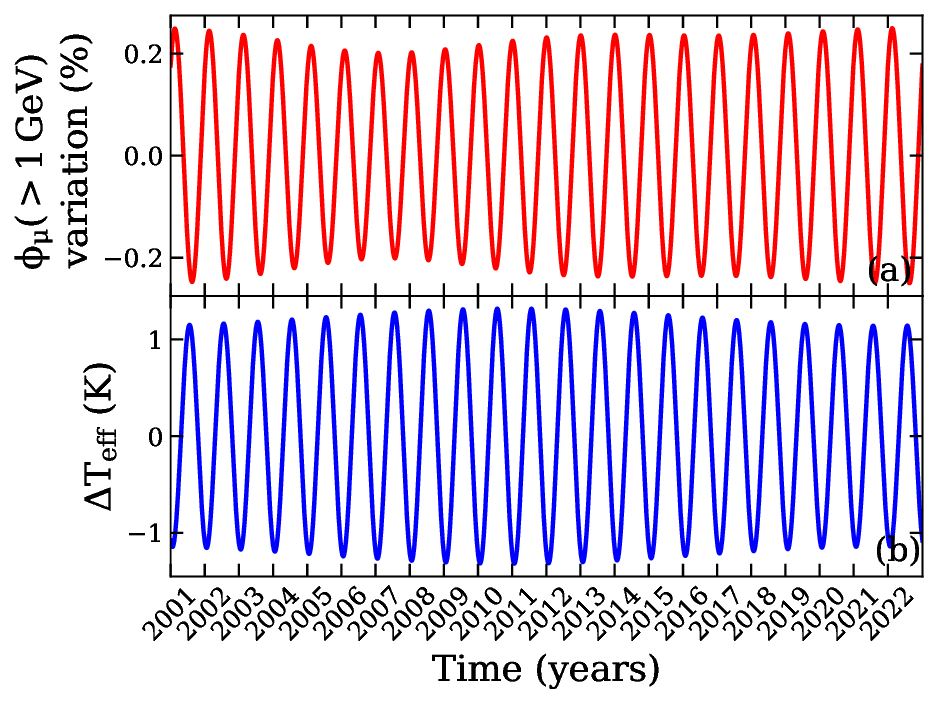}
    \caption{Inverse fast Fourier transform (IFFT) data in the time domain for (a) muon flux variation (in $\%$) and (b) $\Delta{\text{T}_{\text{eff}}}$ (in K; assuming $\lambda$=120 g\,cm$^{-2}$) over the 22 years (2001--2022).}
    \label{Fig_7}
\end{figure}

Subsequently an inverse fast Fourier transform (IFFT) is performed on the filtered power spectrum spectrum to obtain the data in the time domain. The resulting seasonal components of muon flux variation and temperature deviation are shown in Figure \ref{Fig_7}. A small amplitude of approximately $0.2\%$ in the muon flux variation and about 1 K in the temperature deviation are observed.

\begin{figure}
    \centering
    \includegraphics[width=0.48\textwidth]{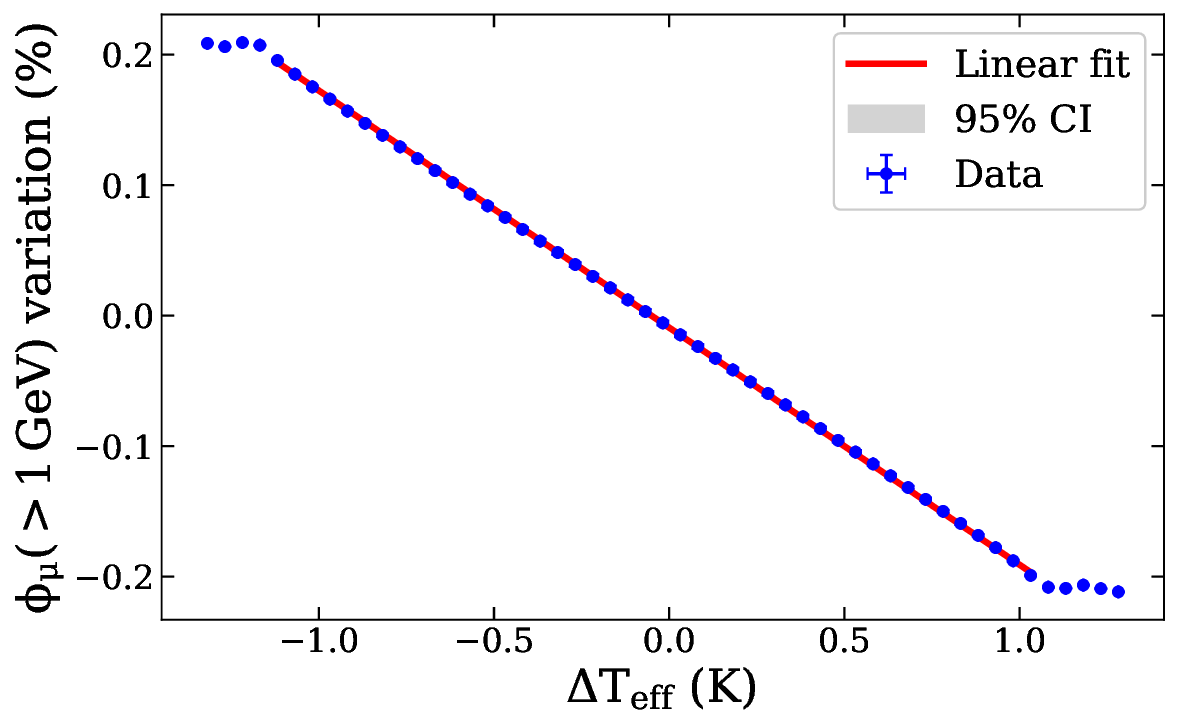}
    \caption{Variation of muon flux (in $\%$) as a function of $\Delta{\text{T}_{\text{eff}}}$ (in K; assuming $\lambda$=120 g\,cm$^{-2}$). The statistical uncertainties are smaller than the plotting symbols and are therefore not visible. {\color{black}The solid red line represents the linear fit, while the grey shaded region indicates the 95$\%$  confidence interval (CI) of the fit, which is very small and largely overlaps the fitted line.} The slope of the linear regression yields a temperature coefficient, $\alpha_{\text{T}}=-\,0.1817\,\pm\,0.0002\,\%\,\text{K}^{-1}$.}
    \label{Fig_8}
\end{figure}

To determine $\alpha_\text{T}$, the filtered muon flux variation is plotted as a function of the corresponding filtered temperature deviation, as shown in Figure \ref{Fig_8}. Each point represents the mean change in muon flux within 0.05 K bins of $\Delta {\text{T}_{\text{eff}}}$. A linear fit to this data, represented by the red solid line in Figure \ref{Fig_8} yields a slope of $-0.1817\pm 0.0002 \, \% \text{K}^{-1}$ representing the temperature coefficient ($\alpha_{\text{T}}$).

\subsection{The magnetic field coefficient ($\gamma_{\text{M}}$)}
\label{section:gamma_M}

The influence of the interplanetary magnetic field (IMF) on the muon flux observed by the GRAPES-3 telescope is investigated after removing the temperature-dependent variations, thereby isolating the heliospheric modulation from atmospheric effects. The temperature coefficient, determined in the previous section to be $\alpha_{\text{T}}$= $- 0.1817\,\%\, \text{K}^{-1}$ for a hadronic attenuation length $\lambda=120\,\text{g cm}^{-2}$, is employed to correct the observed muon flux for atmospheric temperature according to the relation:
\begin{equation}
    \text{R}_{\text{cor}}=\frac{\text{R}_{\text{obs}}}{1+\alpha_{\text{T}} \Delta \text{T}_{\text{eff}}},\label{eq:T_cor}
\end{equation}
where $\text{R}_{\text{obs}}$ is the measured muon rate, $\Delta \text{T}_{\text{eff}}$ is the deviation of the effective atmospheric temperature from its mean, and $\text{R}_{\text{cor}}$ denotes the temperature-corrected muon rate. This ensures that the remaining variations primarily reflect solar and heliospheric influences.

\begin{figure}
    \centering
    \includegraphics[width=0.48\textwidth]{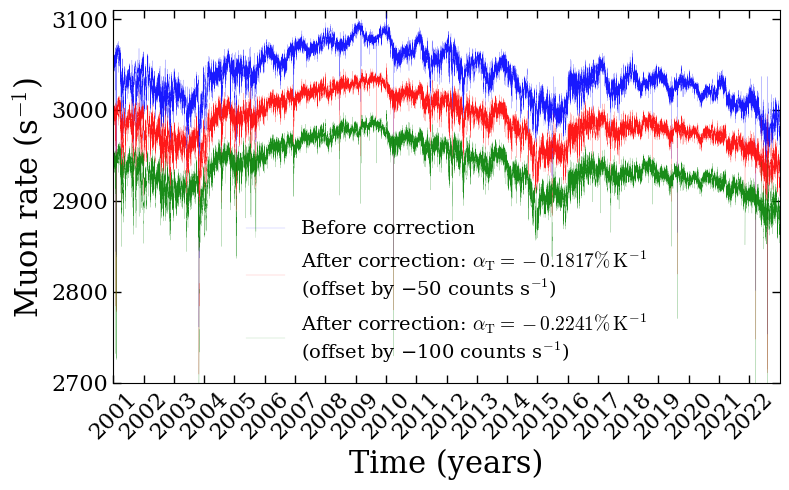}
    \caption{Temporal variation of the muon rate before (blue) and after (red) correcting for the effect of atmospheric temperature using $\alpha_\text{T}$ = $-0.1817\, \% \text{K}^{-1}$ (assuming $\lambda$=120 g\,cm$^{-2}$) over the 22 years (2001--2022). For visual clarity, the temperature-corrected muon rate has been shifted downward by 50 counts s$^{-1}$. For comparison, the muon rate corrected using the final value of $\alpha_\text{T} = -0.2241\, \% \text{K}^{-1}$ is also depicted in green, shifted downward by 100 counts s$^{-1}$.}
    \label{Fig_10}
\end{figure}

Figure \ref{Fig_10} shows the muon rate before (blue) and after (red) applying the temperature correction over 22 years (2001–2022). The temperature-corrected data reveal the long-term modulation of cosmic-ray muon flux that was previously coupled with upper atmospheric temperature variations. The corrected muon rate shows a clear $\approx$11-year modulation driven by long-term variations in the IMF.

\begin{figure}
    \centering
    \includegraphics[width=0.48\textwidth]{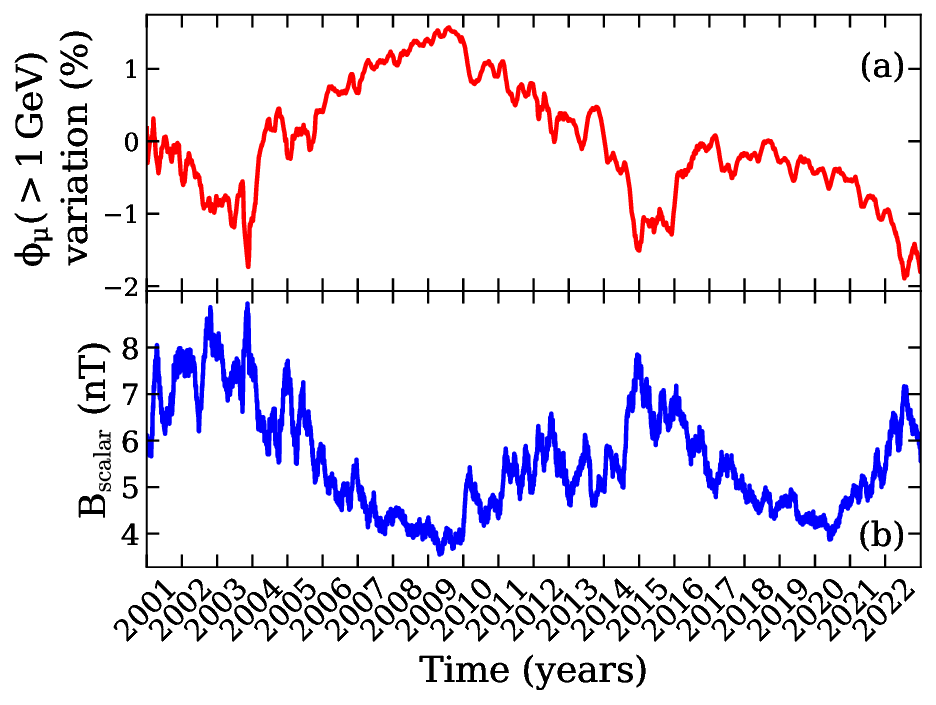}
    \caption{ (a) Percentage variation in the temperature-corrected (using $\alpha_\text{T}$ = $-0.1817\, \% \text{K}^{-1}$) muon flux , and (b) $\text{B}_{\text{scalar}}$ (nT), over the period 2001--2022, after a 60-day running average low-pass filter has been applied.}
    \label{Fig_11}
\end{figure}

Figures \ref{Fig_11}(a) and \ref{Fig_11}(b) depict the resultant percentage deviations in the temperature-corrected muon flux and the IMF (B$_{\text{scalar}}$) after applying the 60 day running average low pass filter. A pronounced anti-correlation with an $\approx$11-year periodicity is evident: higher IMF strengths correspond to reduced muon flux, consistent with the expected suppression of GCRs during periods of enhanced heliospheric magnetic field intensity.

To quantify this relationship, we model the fractional deviation in the muon flux as follows:
\begin{equation}
    \frac{\text{R}-\Bar{\text{R}}}{\Bar{\text{R}}}\times 100(\%)=\gamma_{\text{M}}\times \Delta{\text{B}_{\text{scalar}}} \label{eq:gamma_M}
\end{equation}
where R is the temperature-corrected muon flux at a given time, $\Bar{\text{R}}$ represents the mean, and $\Delta{\text{B}_{\text{scalar}}}$ = $\text{B}_{\text{scalar}}-\Bar{\text{B}}_{\text{scalar}}$ is the deviation of the IMF from its mean value. The magnetic field coefficient, `$\gamma_{\text{M}}$' (expressed in $\%\, \text{nT}^{-1}$), quantifies the fractional change in muon flux per unit change in IMF strength. 

\begin{figure}
    \centering
    \includegraphics[width=0.48\textwidth]{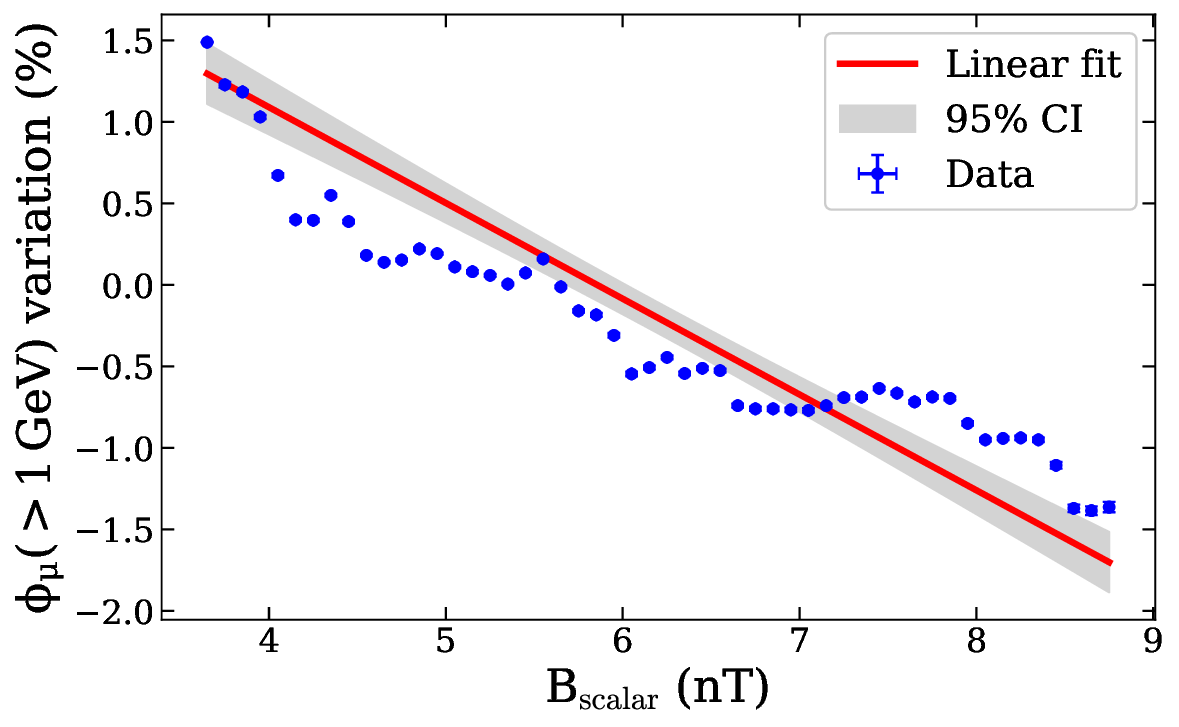}
    \caption{Variation of the temperature-corrected (using $\alpha_{\text{T}}=-\,0.1817\, \% \text{K}^{-1}$) muon flux  as a function of $\text{B}
    _\text{scalar}$ (nT). Each point represents the mean percent deviation of the temperature-corrected muon rate for a 0.01 nT change in B$_{\text{scalar}}$. B$_{\text{scalar}}$. The statistical uncertainties are smaller than the plotting symbols and are therefore not depicted. {\color{black}The solid red line represents the linear fit, while the grey shaded region indicates the 95$\%$ CI of the fit.} The linear regression yields a magnetic field coefficient, $\gamma_{\text{M}}=-\,0.587\,\pm\,0.026\,\% \text{nT}^{-1}$.}
    
    \label{Fig_12}
\end{figure}

To determine $\gamma_\text{M}$, the variation in the temperature-corrected muon flux is plotted against the corresponding IMF magnitude as shown in Figure \ref{Fig_12}. A linear fit to the data (solid red line in Figure \ref{Fig_12}) yields a magnetic field coefficient $\gamma_{\text{M}}$ = $- 0.587 \pm 0.026 \, \% \text{nT}^{-1}$.

\subsection{Iterative refinement}

\begin{figure}
    \centering
    \includegraphics[width=0.48\textwidth]{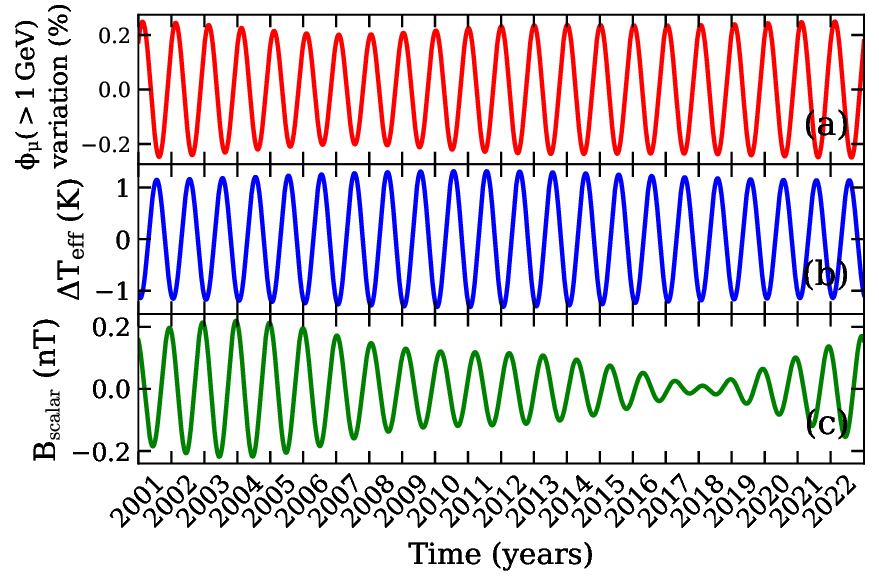}
    \caption{Inverse fast Fourier transform (IFFT) data (using the band-pass filter $W(f)$, Equation \ref{eq:band-pass}) in the time domain for (a) muon flux variation (in $\%$), (b) $\Delta{\text{T}_{\text{eff}}}$ (in K; assuming $\lambda$=120 g\,cm$^{-2}$), and (c) $\text{B}_{\text{scalar}}$ (in nT) over the 22 years (2001--2022).} 
    \label{Fig_93}
\end{figure}

Although the band pass filter effectively isolates the seasonal modulation of muon flux caused by atmospheric temperature variations from other periodicities, it still allows the effects of IMF modulations with a similar periodicity to that of the atmospheric temperature to contaminate the temperature dependence. The presence of concomitant modulations in the $\text{B}_{\text{scalar}}$ component, even with an amplitude as small as 0.2 nT (see Figure \ref{Fig_93}(c)), has a nontrivial impact on the estimation of $\alpha_{\text{T}}$. To mitigate this, we employ an iterative fitting process that alternately removes the effects of IMF and temperature variations from the muon data while fitting for the other.

\begin{figure}
    \centering
    \includegraphics[width=0.48\textwidth]{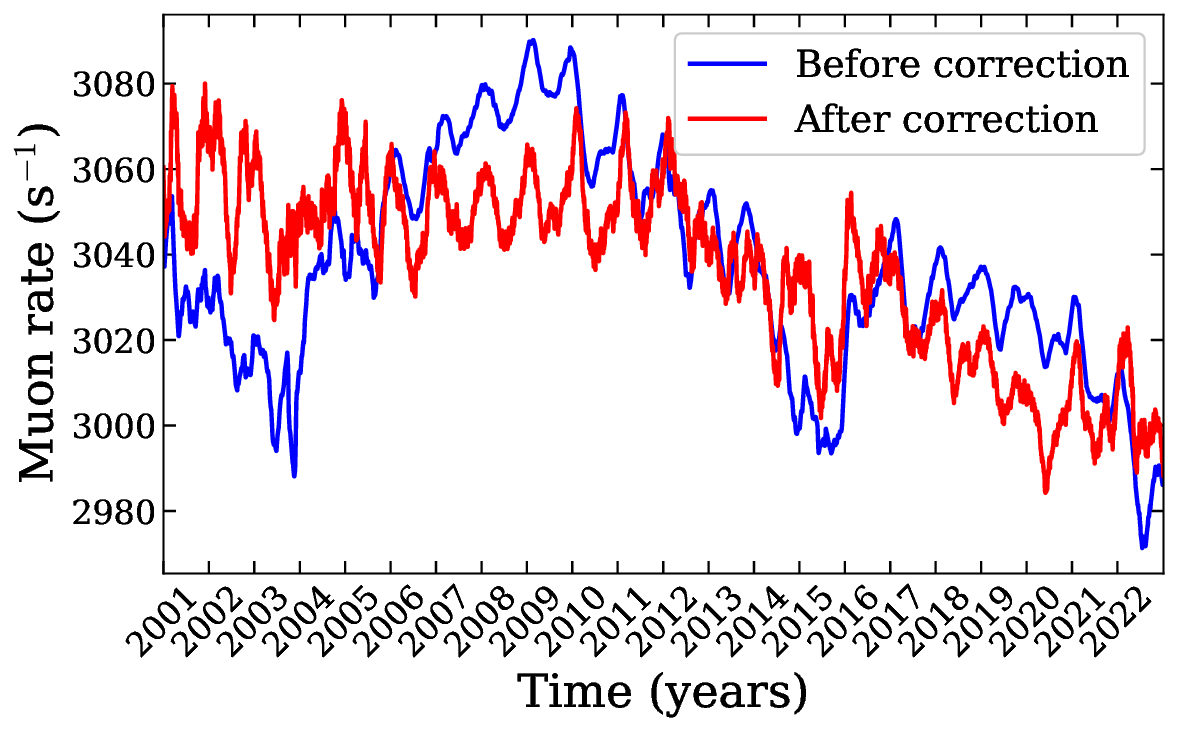}
    \caption{Temporal variation of the muon rate before (blue) and after (red) correcting for the effects of the IMF using $\gamma_\text{M}$ = $-0.587\, \% \, \text{nT}^{-1}$ (assuming $\lambda$=120 g\,cm$^{-2}$), both obtained using a 60-day low-pass filter over the 22 years (2001--2022).}
    \label{Fig_13}
\end{figure}

\begin{figure}
    \centering
    \includegraphics[width=0.48\textwidth]{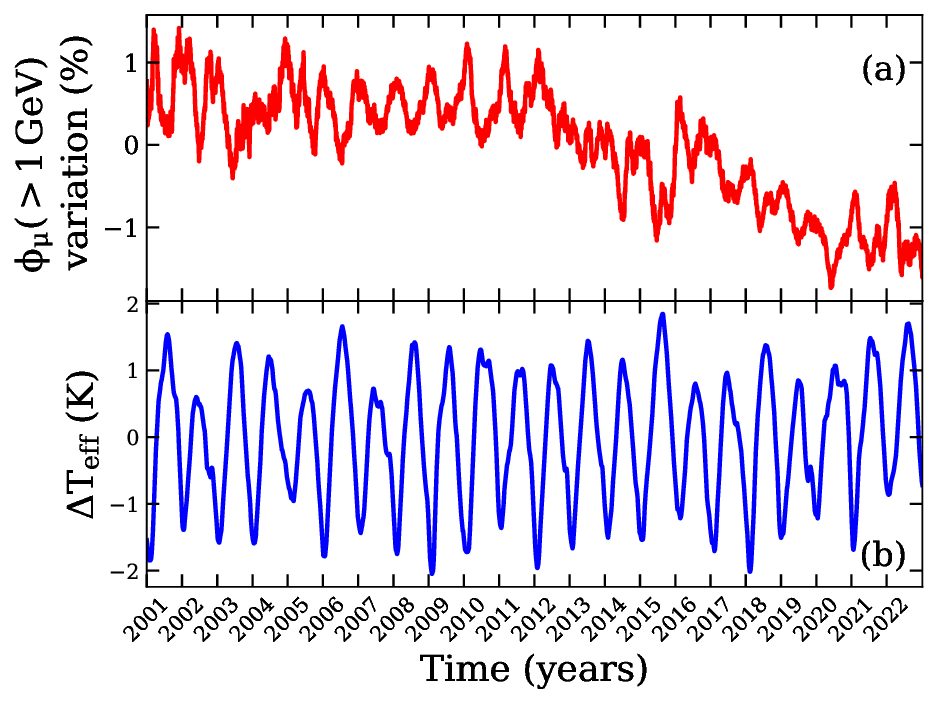}
    \caption{Variation of (a) magnetic field-corrected muon flux (in $\%$; using $\gamma_\text{M}$ = $-0.587\, \% \, \text{nT}^{-1}$), and (b) $\Delta{\text{T}_{\text{eff}}}$ (in K; assuming $\lambda$=120 g\,cm$^{-2}$), obtained using a 60-day low-pass filter over the period (2001--2022).}
    \label{Fig_14}
\end{figure}

In the first iteration, the muon rate is initially corrected for the effect of the IMF using the previously determined magnetic field coefficient, $\gamma_{\text{M}}$= $-0.587\, \% \, \text{nT}^{-1}$ using the following equation:
\begin{equation}
    \text{R}_{\text{cor}}=\frac{\text{R}_{\text{obs}}}{1+\gamma_{\text{M}} \Delta \text{B}_{\text{scalar}}},\label{eq:M_cor}
\end{equation}
where $\text{R}_{\text{obs}}$ is the observed muon rate, $\Delta \text{B}_{\text{scalar}}$ is the deviation of the IMF from its mean, and $\text{R}_{\text{cor}}$ is the IMF-corrected muon rate. Figure \ref{Fig_13} compares the uncorrected muon rates (blue) and the magnetic field corrected muon rates (red), over the 22 years (2001--2022) after the 60 day running average has been applied. The IMF correction effectively suppresses the slow, 11-year solar modulation component, thereby enhancing the visibility of the annual variation in muon flux driven by atmospheric temperature changes. The known anti-correlation between the muon flux variations and temperature deviations becomes more pronounced after removing the IMF influence, as seen in figures \ref{Fig_14}(a) and \ref{Fig_14}(b).

\begin{figure}
    \centering
    \includegraphics[width=0.48\textwidth]{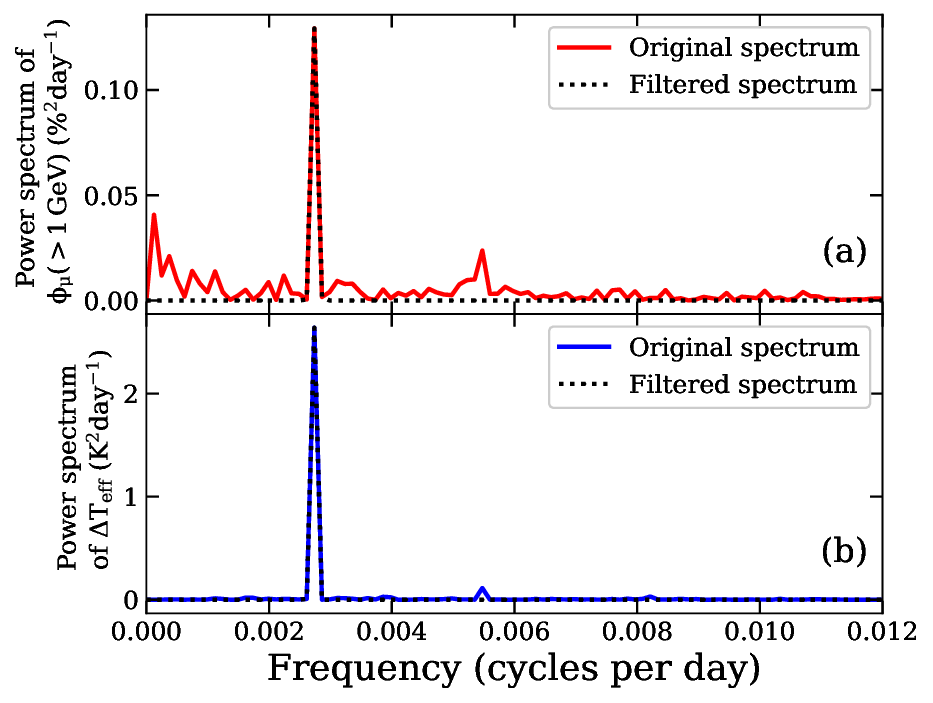}
    \caption{Fast Fourier transform (FFT) spectra of (a) magnetic field-corrected muon flux variation (in $\%$; using $\gamma_\text{M}$ = $-0.587\, \% \, \text{nT}^{-1}$) and (b) $\Delta{\text{T}_{\text{eff}}}$ (in K; assuming $\lambda$=120 g\,cm$^{-2}$) over the 22 years (2001--2022).}
    \label{Fig_15}
\end{figure}

Following this IMF correction, the FFT analysis is repeated on the corrected muon data together with the effective temperature series. Figure \ref{Fig_15}(a) presents the FFT spectrum of the IMF-corrected muon flux, while Figure \ref{Fig_15}(b) shows that of the temperature deviation $\Delta \text{T}_{\text{eff}}$ (blue). The amplitude of the annual component is significantly enhanced, while the long-term 11-year solar modulation peak is strongly reduced, consistent with expectations of the iterative correction effectively isolating the atmospheric contribution.

\begin{figure}
    \centering
    \includegraphics[width=0.48\textwidth]{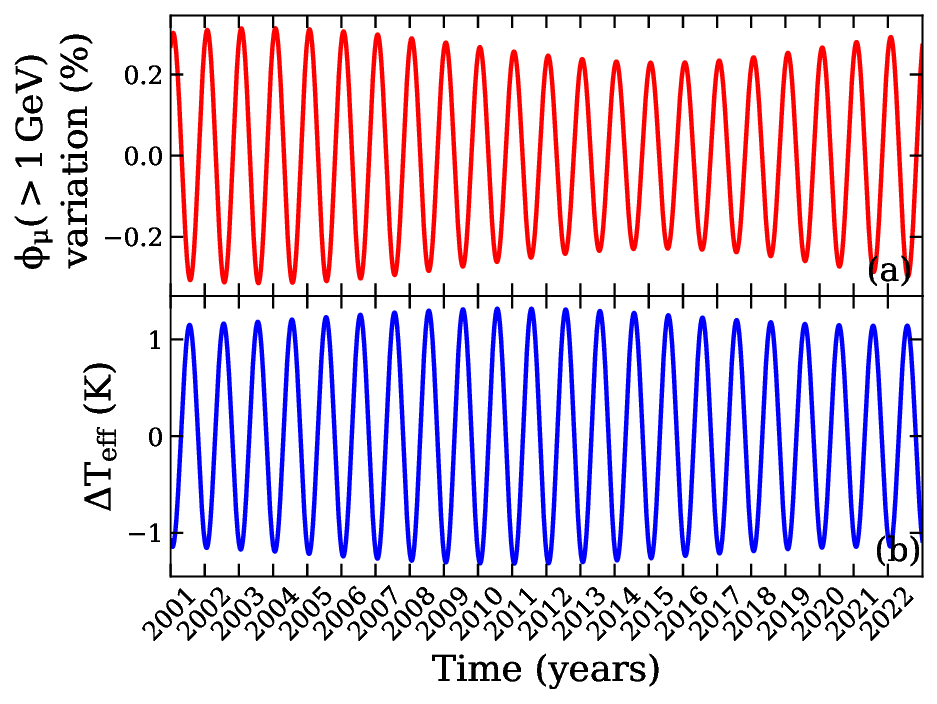}
    \caption{Band pass filtered data for (a) magnetic field-corrected muon flux variation (in $\%$; using $\gamma_\text{M}$ = $-0.587\, \% \, \text{nT}^{-1}$) and (b) $\Delta{\text{T}_{\text{eff}}}$ (in K; assuming $\lambda$=120 g\,cm$^{-2}$) over the 22 years (2001--2022).}
    \label{Fig_16}
\end{figure}

To further isolate the annual periodicity, the same band-pass filter $W(f)$ (defined in Equation \ref{eq:band-pass}) was applied to both datasets. The resulting filtered spectra, shown as black dotted lines in Figure \ref{Fig_15}, retain only the annual component associated with atmospheric temperature modulation. An IFFT was then performed on the filtered spectra to reconstruct the data in the time domain. The reconstructed IFFT time series for the IMF-corrected muon flux (red) and the effective temperature deviation (blue) are displayed in Figures \ref{Fig_16}(a) and \ref{Fig_16}(b), respectively. A clear anti-correlation between the two time series is evident.

\begin{figure}
    \centering
    \includegraphics[width=0.48\textwidth]{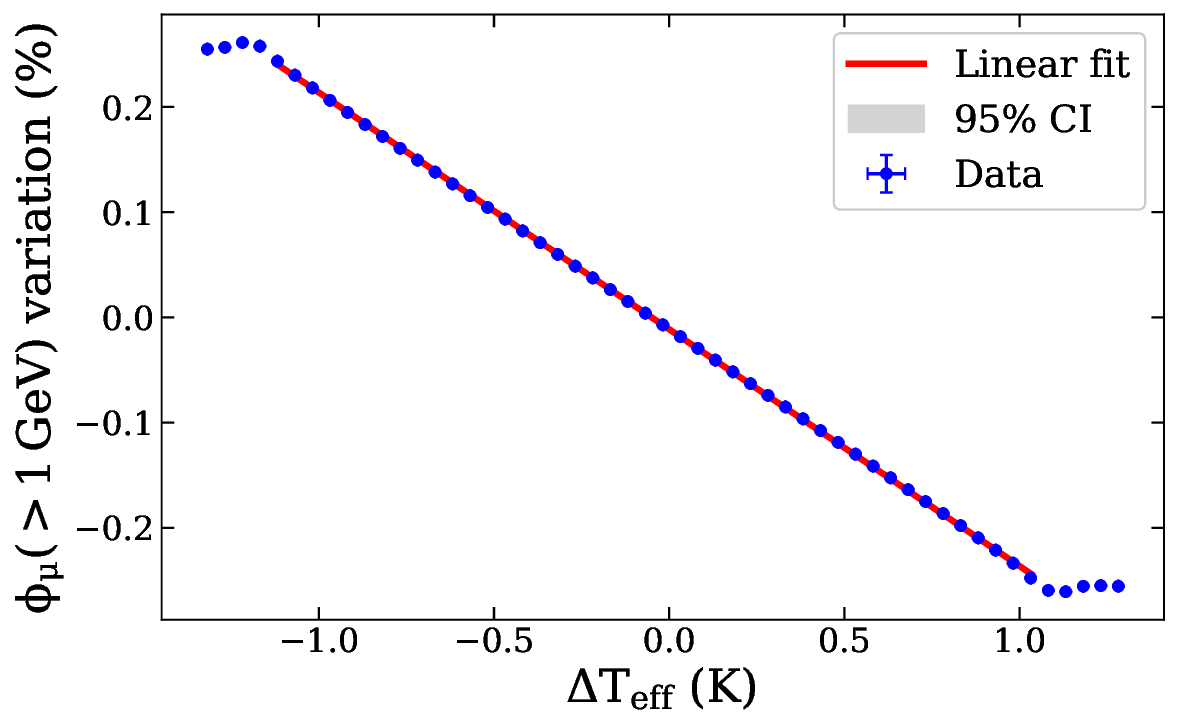}
    \caption{Variation of magnetic field-corrected muon flux (in $\%$; using $\gamma_\text{M}$ = $-0.587\, \% \, \text{nT}^{-1}$) as a function of $\Delta{\text{T}_{\text{eff}}}$ (in K; assuming $\lambda$=120 g\,cm$^{-2}$). {\color{black}The solid red line represents the linear fit, while the grey shaded region indicates the 95$\%$ CI of the fit, which is very small and largely overlaps the fitted line.} The linear regression yields a temperature coefficient, $\alpha_{\text{T}}=-\,0.2251\,\pm\,0.0003\,\%\,\text{K}^{-1}$.}
    \label{Fig_17}
\end{figure}

To determine $\alpha_\text{T}$, the filtered IMF-corrected muon flux variation was plotted as a function of the corresponding filtered temperature deviation, as shown in Figure \ref{Fig_17}. Each data point corresponds to the mean muon flux change within a 0.05 K temperature bin. Statistical uncertainties are smaller than the plotting symbols and are therefore not visible. A linear fit to the data, shown as the red solid line in Figure \ref{Fig_17}, gives $\alpha_{\text{T}} = -0.2251 \pm 0.0003\,\%\,\text{K}^{-1}$, after IMF correction. As discussed earlier, the use of FFT and filtering significantly reduces non-atmospheric noise, leading to a small statistical uncertainty.

\begin{figure}
    \centering
    \includegraphics[width=0.48\textwidth]{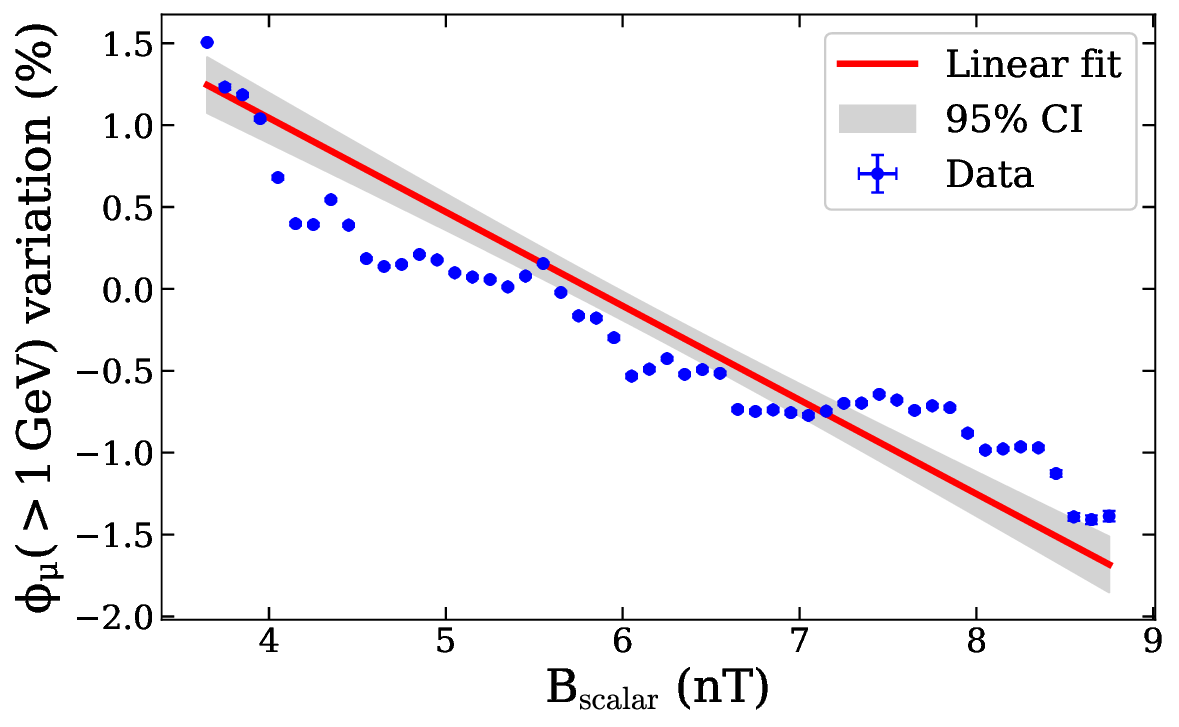}
    \caption{Variation of temperature-corrected muon flux (in $\%$; using $\alpha_{\text{T}}=-\,0.2251\, \% \text{K}^{-1}$) as a function of $\text{B}
    _\text{scalar}$ (nT). {\color{black}The solid red line represents the linear fit, while the grey shaded region indicates the 95$\%$ confidence interval (CI) of the fit.} The linear regression yields a magnetic field coefficient, $\gamma_{\text{M}}=-\,0.574\,\pm\,0.027\,\% \text{nT}^{-1}$.}

    \label{Fig_18}
\end{figure}

This refined $\alpha_{\text{T}}$ value was then used to correct the muon rate for temperature effects, after which the magnetic field coefficient $\gamma_{\text{M}}$ was re-evaluated as described in Section \ref{section:gamma_M}. Figure \ref{Fig_18} shows variation of temperature corrected muon flux ( using $\alpha_{\text{T}}=-\,0.2251\, \% \text{K}^{-1}$) as a function of $\text{B}_\text{scalar}$ (nT). Each point represents the mean percentage deviation of the muon rate in 0.01 nT bins, with uncertainties depicted as vertical and horizontal error bars. The linear fit (solid red line) yields a magnetic field coefficient, $\gamma_{\text{M}}=-\,0.574\,\pm\,0.027\,\% \, \text{nT}^{-1}$. Thus, after the first iteration, the temperature and magnetic field coefficients were found to be $\alpha_{\text{T}}$ = $-0.2251 \pm 0.0003 \, \% \,\text{K}^{-1}$ and $\gamma_{\text{M}}$ = $-0.574 \pm 0.027 \, \% \,\text{nT}^{-1}$, respectively.

\begin{figure}
    \centering
    \includegraphics[width=0.48\textwidth]{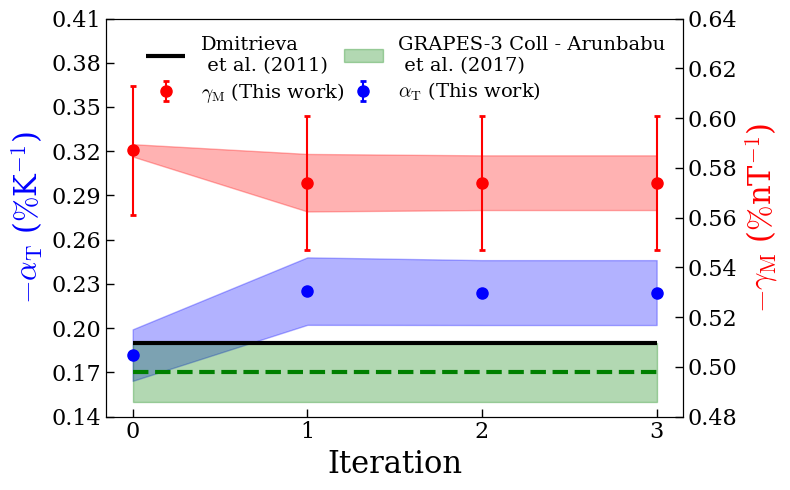}
    \caption{ Progression of $\alpha _{\text{T}}$ (blue dots) and $\gamma_{\text{M}}$ (red dots) for an assumed hadronic attenuation length $\lambda$=120 g cm$^{-2}$ through successive iterations. The vertical error bars indicate statistical uncertainties, while the surrounding blue and red shaded bands denote the corresponding systematic uncertainties originating from uncertainty in $\lambda$. The green dashed horizontal line and its surrounding shaded band show $\alpha _{\text{T}}$ and its uncertainty previously reported by the GRAPES-3 Collaboration ~\cite{2017APh....94...22A} based on 6 years of data (2005-2010), and the black horizontal line indicates the central value from theoretical predictions~\cite{2011APh....34..401D}.}
    \label{Fig_21}
\end{figure}

 The iterative procedure is then repeated and it was observed that $\alpha_{\text{T}}$ and $\gamma_{\text{M}}$ values stabilize quickly (see Fig.~\ref{Fig_21}). .The dependence of $\alpha_{\text{T}}$ and $\gamma_{\text{M}}$ on $\lambda$ was examined in a range of 80 to 180 g cm$^{-2}$ in steps of 20 g cm$^{-2}$. Table \ref{table5} summarizes the results at the final iteration. After three iterations, both $\alpha_{\text{T}}$ and $\gamma_{\text{M}}$ converge for all values of $\lambda$.

\begin{figure}
    \centering
    \includegraphics[width=0.48\textwidth]{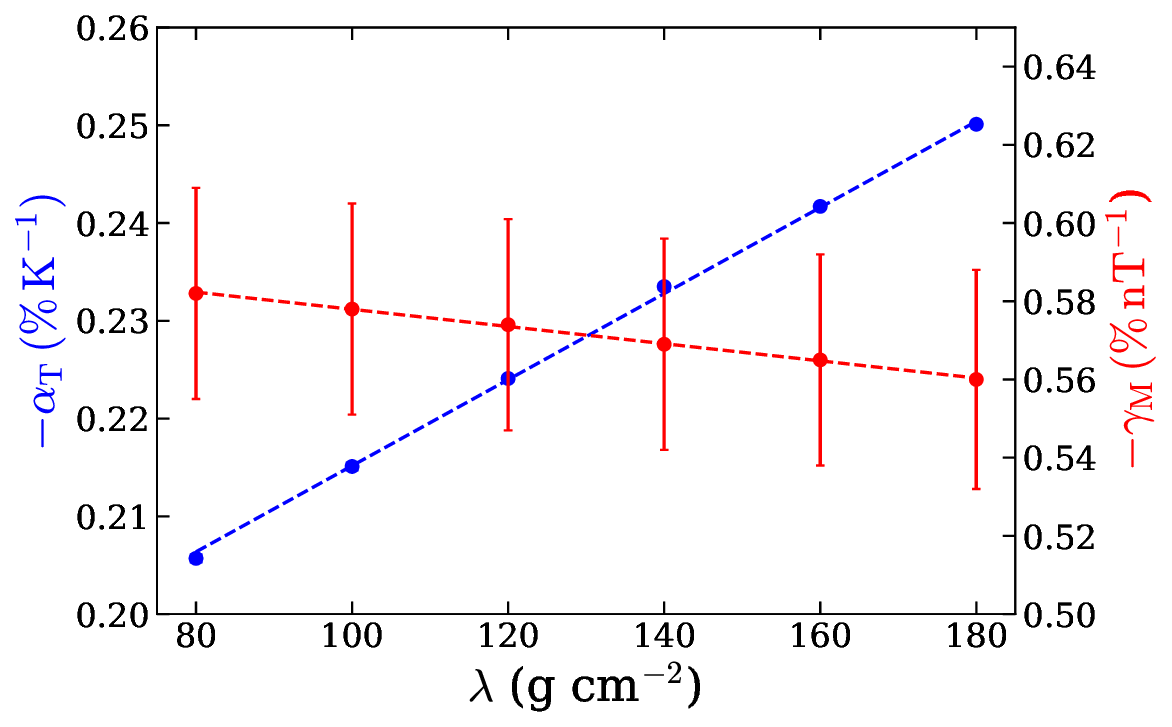}
    \caption{Dependence of the final (Iteration 3) $\alpha _{\text{T}}$ and $\gamma_{\text{M}}$   on the assumed hadronic attenuation length ($\lambda$). The dashed lines show the corresponding linear fits and the vertical error bars indicating statistical uncertainties.}
    \label{Fig_20}
\end{figure}

To estimate the systematic uncertainty associated with the choice of $\lambda$, the variation of $\alpha _{\text{T}}$ and $\gamma_{\text{M}}$ with $\lambda$ was examined at each iteration. For every iteration, the dependence of both coefficients on $\lambda$ was approximated by a linear fit to quantify their sensitivity to the assumed attenuation length. As shown in Figure \ref{Fig_20} (for iteration 3 as a representative case) $\alpha _{\text{T}}$ exhibits a gradual increase in magnitude with $\lambda$, whereas $\gamma_{\text{M}}$ shows a weakly decreasing trend. The dashed lines represent the corresponding linear fits, and the vertical error bars indicate statistical uncertainties. The systematic uncertainty in each iteration was derived from the product of the fitted slope and $\Delta\lambda$, where $\Delta\lambda$ = 50 g cm$^{-2}$ (half the studied $\lambda$ range), representing the response of each coefficient to uncertainty in the hadronic attenuation length. Figure \ref{Fig_21} presents the progression of $\alpha _{\text{T}}$ (blue dots) and mean $\gamma_{\text{M}}$ (red dots) for an assumed hadronic attenuation length $\lambda$=120 g cm$^{-2}$ through successive iterations. The vertical error bars indicate statistical uncertainties, while the surrounding blue and red shaded bands denote the corresponding systematic uncertainty ranges. For iteration 0, the temperature coefficient was determined to be $\alpha_{\text{T}}=-\,0.1817\,\pm\,0.0002\, (\text{stat.})\,\pm\,0.0175\, (\text{syst.})\,\%\,\text{K}^{-1}$, and the magnetic field coefficient as $\gamma_{\text{M}}=-\,0.587\,\pm\,0.026\, (\text{stat.})\,\pm\,0.002\, (\text{syst.})\,\%\,\text{nT}^{-1}$. These results are consistent with earlier studies, which reported $\alpha_{\text{T}}$= -- 0.17 $\pm$ 0.02 $\%\, \text{K}^{-1}$ \cite{2017APh....94...22A} and $\alpha_{\text{T}}$= -- 0.19 $\% \,\text{K}^{-1}$ \cite{2011APh....34..401D}. After the third iteration, the coefficients converged to $\alpha_{\text{T}}=-\,0.2241\,\pm\,0.0003\, (\text{stat.})\,\pm\,0.0220\, (\text{syst.})\,\%\,\text{K}^{-1}$ and $\gamma_{\text{M}}=-\,0.574\,\pm\,0.027\, (\text{stat.})\,\pm\,0.011\, (\text{syst.})\,\%\,\text{nT}^{-1}$, confirming the robustness and self-consistency of the iterative fitting method. {\color{black}Propagating the uncertainty of the MERRA-2 temperature data ($0.962\,\mathrm{K}$) through the effective temperature calculation results in an additional statistical uncertainty of $0.04\,\%\,\mathrm{K^{-1}}$ in the temperature coefficient $\alpha_\mathrm{T}$.}

\begin{table}
\centering
\caption{Dependence of $\alpha_{\text{T}}$ and $\gamma_{\text{M}}$ on $\lambda$ for iteration 0.}
\begin{tabular}{lccc} 
\hline
$\lambda \, ({\text{g\,cm}}^{-2})$ & $\alpha_{\text{T}}\, (\%\,\text{K}^{-1})$ & $\gamma_{\text{M}} (\%\,\text{nT}^{-1})$ \\ 
\hline
80  & $-0.1671 \pm 0.0003$ &  $-0.588\pm 0.026$ \\ 
100 & $-0.1749 \pm 0.0003$ &  $-0.588\pm 0.026$ \\
120 & $-0.1817 \pm 0.0002$  & $-0.587\pm 0.026$\\
140 & $-0.1893 \pm 0.0002$  & $-0.586\pm 0.026$\\
160 & $-0.1957 \pm 0.0001$  & $-0.585\pm0.027$\\
180 & $-0.2026 \pm 0.0001$  & $-0.583\pm0.027$\\
\hline
\end{tabular}
\label{table2}
\end{table}

\begin{table} 
\centering
\caption{Dependence of $\alpha_{\text{T}}$ and $\gamma_{\text{M}}$ on $\lambda$ for iteration 3.}
\begin{tabular}{lccc}
\hline
$\lambda \, ({\text{g\,cm}}^{-2})$ & $\alpha_{\text{T}}\, (\%\,\text{K}^{-1})$ & $\gamma_{\text{M}} (\%\,\text{nT}^{-1})$ \\
\hline
80  & $-0.2057\pm0.0004$ & $-0.582\pm0.027$ \\
100 & $-0.2151\pm0.0004$ & $-0.578\pm0.027$ \\
120 &  $-0.2241\pm0.0003$& $-0.574\pm0.027$ \\
140 & $-0.2335\pm0.0003$ & $-0.569\pm0.027$ \\
160 & $-0.2417\pm0.0002$ & $-0.565\pm0.027$ \\
180 & $-0.2501\pm0.0002$& $-0.560\pm0.028$  \\
\hline
\end{tabular}

\label{table5}
\end{table}

\section{Summary of results}
The GRAPES-3 muon telescope, with its 16 independent modules, has provided an uninterrupted high-statistics record of the atmospheric muon flux over 22 years (2001--2022), covering parts of three solar cycles partially. Novel algorithmic methods~\cite{Paul:2025iet} have enabled the mitigation of detector instabilities and efficiency variations, by exploiting the high level of redundancy offered by the 16 independent modules of G3MT. The resultant long duration dataset enabled a detailed investigation of how variations in upper atmospheric temperature and the interplanetary magnetic field influence ground-level muon flux. 

After applying a 60-day running average low-pass filter, an analysis using FFT and a band pass filter returns $\alpha_{\text{T}}=-\,0.1817\,\pm\,0.0002 (\text{stat.})\,\%\,\text{K}^{-1}$ and $\gamma_{\text{M}}=-\,0.587\,\pm\,0.027\,(\text{stat.})\,\%\,\text{nT}^{-1}$, compatible with what was reported earlier by the GRAPES-3 Collaboration~\cite{2017APh....94...22A} from six years (2005-2010) of data; $\alpha_{\text{T}}=-\,0.17\,\pm\,0.02\%\,\text{K}^{-1}$. An iterative deconvolution leads to a $\sim 23\%$ increase in temperature coefficient, and a concomitant $\sim 2.3\%$ reduction in the magnetic field coefficient; $\alpha_{\text{T}}=-\,0.2241\,\pm\,0.0003\, (\text{stat.})\,\%\,\text{K}^{-1}$ and $\gamma_{\text{M}}=-\,0.574\,\pm\,0.027\, (\text{stat.})\,\%\,\text{nT}^{-1}$. The principal systematic uncertainty on both quantities was found to be the assumed hadronic attenuation length $\lambda$. For $\Delta\lambda$ = 50 g cm$^{-2}$, the systematic uncertainties on $\alpha_{\text{T}}$ and $\gamma_{\text{M}}$ are $\pm 0.0220\, \%\,\text{K}^{-1}$ and $\pm 0.011\, \,\%\,\text{nT}^{-1}$ respectively.

These results demonstrate the potential of the GRAPES-3 muon telescope to serve as a long term monitor of both the upper atmospheric temperature and the interplanetary magnetic field to within 20\% and 6\% respectively. 

\section{Discussion \& conclusion}

The temperature dependence of the all sky cumulative muon flux recorded by G3MT over 22 years has been found to be compatible with what was reported previously from 6 years of G3MT data~\cite{2017APh....94...22A} and an iterative method deconvolutes the effects of the modulation of the primary CR flux by the IMF to provide a more refined value of $\alpha_{\mathrm{T}}$. The results are in the range reported by other comparable experiments~\cite{https://doi.org/10.1029/2020EA001131, 2023IJAA...13..236M, gi-10-219-2021, Savic:2021Wb,2017arXiv170104030B} and relevant simulation studies~\cite{2011APh....34..401D}. This serves as both a validation of the methods used to mitigate long term variations in the efficiency of G3MT modules as well as an accentuation of the potential of atmospheric CR secondary flux measurements to serve as live monitors of the conditions of the upper atmosphere, in the presence of independent data about the IMF. With the advent of the Aditya-L1 mission~\cite{2017CSci..113..610S}, and the availability of near real time data on the IMF, it becomes imperative to characterize these relationships for each of the 169 independent directions of G3MT to enable temperature imaging of the upper atmosphere. The potential of these data to improve the deterministic predictability of long term Earth climate models using ensemble forecasting techniques remains to be explored~\cite{Palmer2019}.

\section*{Acknowledgements}

We thank D.B. Arjunan, Late A.A. Basha, G.P. Francis, I.M. Haroon, V. Jeyakumar, Late S. Karthikeyan, S. Kingston,
N.K. Lokre, S. Murugapandian, S. Pandurangan, P.S. Rakshe, K. Ramadass, C. Ravindran, K.C. Ravindran, V. Santosh Kumar, S. Sathyaraj, M.S. Shareef, C. Shobana, R. Suresh Kumar, K. Viswanathan, and V. Viswanathan for their assistance over the years in running the experiment. We acknowledge the support of the Department of Atomic Energy, Government of India, under Project Identification No. RTI4002.

\bibliographystyle{elsarticle-num}
\bibliography{bibfile}

\end{document}